\documentclass[prb,12pt]{revtex4}

\usepackage{amsmath}    
\usepackage{graphicx}   
\usepackage{verbatim}   
\usepackage{color}      
\usepackage{subfigure}  
\usepackage{hyperref}   
\usepackage{epstopdf}

\def\blue{\color{black}}
\begin{document}

\title{Chameleon Gravity, Electrostatics, and Kinematics in the Outer Galaxy}

\author{R. Pourhasan$^{(a)}$\footnote{email address:
r2pourha@uwaterloo.ca}, N. Afshordi$^{(b, a)}$\footnote{email
address: nafshordi@perimeterinstitute.ca}, R. B.
Mann$^{(a,b)}$\footnote{email address: rbmann@sciborg.uwaterloo.ca}
and A.~C. Davis$^{(c)}$\footnote{email address:
A.C.Davis@damtp.cam.ac.uk}} \affiliation{$^{(a)}$ Department of
Physics \& Astronomy, University of
Waterloo, Waterloo, Ontario N2L 3G1, Canada \\
$^{(b)}$ Perimeter Institute for Theoretical Physics, 31 Caroline St. N., Waterloo, ON, N2L 2Y5, Canada \\
$^{(c)}$ Department of Applied Mathematics and Theoretical
Physics, Center for Mathematical Sciences, Cambridge CB3 0WA,
United Kingdom}

\begin{abstract}
Light scalar fields are expected to arise in theories of high energy physics (such as string theory), and find phenomenological
motivations in dark energy, dark matter, or neutrino physics.  However, the coupling of light scalar fields to ordinary (or dark) matter
 is strongly constrained from laboratory, solar system, and astrophysical tests of fifth force.  One way to evade these constraints in
 dense environments is through the chameleon mechanism, where the field's mass steeply increases with ambient density. Consequently,
 the chameleonic force is only sourced by a thin shell near the surface of dense objects, which significantly reduces its magnitude.

In this paper, we argue that thin-shell conditions are equivalent to ``conducting'' boundary conditions in electrostatics.
As an application, we use the analogue of the method of images to calculate the back-reaction (or self-force) of an object
around a spherical gravitational source. Using this method, we can explicitly compute the violation of equivalence principle
in the outskirts of galactic haloes (assuming an NFW dark matter profile): Intermediate mass satellites can be slower than
their larger/smaller counterparts by as much as $10\%$ close to a thin shell.

\end{abstract}

\maketitle

\section{Introduction}
Cosmological observations indicate that the universe is accelerating today
with about $75$ percent of the energy density in dark energy
\cite{Perlmutter:1998np, Riess:1998cb}. If the dark
energy is a scalar field then its mass needs to be very light indeed, but its
couplings to ordinary matter must be suppressed to avoid fifth force
constraints. Indeed fifth force experiments such as the Cassini satellite
experiment put stringent bounds on the gravitational coupling of nearly
massless scalar fields \cite{Adelberger:2002ic,Hoskins:1985tn,Decca:2007yb,
Bertotti:2003rm}.  The chameleon scenario posits that a scalar
field with gravitational strength couplings to matter could generate the
present day acceleration, but evade fifth force constraints \cite{Khoury:2003rn}.
The properties of the scalar field depend on the ambient density.
In particular its mass is density dependent.
Cosmologically the chameleon mass can be of order the Hubble constant,
allowing the field to be rolling on cosmological time-scales \cite{Brax:2004qh}. However,
in the solar system the chameleon can be sufficiently massive so that it
evades fifth force constraints. Indeed, chameleon theories have
a non-trivial way of evading empirical gravitational constraints via the
existence of a thin-shell mechanism. For sufficiently large objects the
chameleonic force is almost entirely due to a thin-shell of matter just
below the surface of the object, with the matter in the core of the
object giving a negligible contribution to the force.  Although the original
papers assumed the coupling to matter was of  gravitational strength, it was later
realized that, due to non-linear effects, the coupling could be much
larger whilst still preserving the properties of the chameleon mechanism
\cite{Mota:2006fz}.

The chameleon mechanism has been tested in many different situations
and constraints placed on the parameters of the theory \cite{Brax:2007vm,Brax:2010gp}.
Gravitational tests in the solar system \cite{Khoury:2003rn} and the full
cosmological evolution have been studied \cite{Brax:2004qh}. The effect it has
on structure formation on sub-galactic scales \cite{Brax:2005ew} and on large
scale structure formation have also been investigated \cite{Li:2009sy}.
More recently the
chameleon coupling to photon fields
has been investigated and constraints have been placed on the mechanism from
laboratory experiments \cite{Brax:2007hi} and astrophysical measurements
\cite{Burrage:2008ii}. Laboratory experiments have been designed to look for the
chameleonic force between objects \cite{Brax:2010xx}. These
experiments are similar, though distinct, to Casimir force experiments
and should either detect the chameleonic force or severely constrain
it in the near future.

On cosmological scales, the introduction of a chameleonic force can lead
to violations of the equivalence principle, as objects with shallow potential wells
(or small velocity dispersions) fall at a higher acceleration than the larger object
that have a thin shell \cite{Hui:2009kc}. The chameleon effect on CDM large scale structure
 can now be explicitly seen in numerical simulations of $f(R)$ gravity models
\cite{Oyaizu:2008tb,Li:2011uw,Zhao:2010qy}

One effect that has not been studied is the chameleonic back-reaction
on galactic scale objects. In chameleon theories large galactic
haloes should have a thin shell. In particular, the Navarro-Frenk-White (or NFW) profile models the dark
matter distribution in galactic haloes \cite{Navarro:1996gj}.  In this paper, we investigate the back-reaction
(or self-force) of satellites moving in the halo outskirts of an NFW profile, which can lead to observable violations of the equivalence principle. To do this, we use an analogy between chameleon thin-shell
conditions and electrostatics, which enables an application of the method of images to fifth-force
calculations (This analogy was also recently noticed in \cite{JonesSmith:2011tn}, and used to argue
the presence of `lightning rod' effects close to non-spherical thin shells).

The plan of this paper is as follows: In the next section, we review the
chameleon mechanism, explain how its field depends on the ambient density,
and how the thin shell mechanism works in this context. We then compare the thin shell
mechanism with electrostatics in Sec. \ref{thin_shell}, showing that there is an analogy. This analogy
enables us to use the method of images to compute the chameleonic back-reaction (or self-force). We compute
the self-force corrections to the fifth force between a test object and a body with a thin
shell before going on to consider the NFW profile and
the chameleon force in Sec. \ref{NFW}.   In Sec. \ref{circ_vel}, we show that the circular velocity  of intermediate mass satellites in the outer halo can be reduced significantly if the back-reaction is taken into account.
Finally, Sec. \ref{conclude} contains a discussion of our results and
concludes the paper.

\section{The Chameleon Mechanism}\label{cham_mech}

Chameleon fields appear in scalar-tensor theories of gravity. The action of
the chameleon field in the Einstein frame is
\begin{equation}
S = \int d^4 x\sqrt{-g} \left[\frac{M_p^2}{2}R - \frac{1}{2}(\partial \phi)^2- V(\phi)\right]
\end{equation}
where $M_p=(8\pi G)^{-1/2}$ is the reduced Planck mass.
Matter couples to both gravity and the scalar field according to
\begin{equation}
S_m(\psi, A^2(\phi)g_{\mu\nu}),
\end{equation}
where $\psi$ is a matter field and $A$ the conformal factor relating the
Jordon and Einstein frame metrics. Notice that the scalar field
couples to all matter species, including baryons. This conformal coupling
gives an extra contribution to the Klein-Gordon equation.
\begin{equation}
\nabla^2\phi = V_{,\phi} - \alpha_{\phi}T^{\mu}_{\mu},
\end{equation}
 where $\alpha_{\phi}\equiv\frac{{\partial}lnA}{{\partial}{\phi}}$. In the approximation
that matter is well described by a pressureless, perfect fluid this becomes
\begin{equation}
\nabla^2\phi = V_{,\phi} + \alpha_{\phi}\rho_mA(\phi)
\end{equation}
where $\rho_{m}$ is the matter density.

An immediate consequence is that the dynamics of $\phi$ is governed by
the effective potential, which depends explicitly on the density,
\begin{equation}
V_{eff} = V(\phi) + \rho_m A(\phi).
\end{equation}
If the bare potential is of a runaway form,
\begin{equation}
V(\phi) = {\Lambda^{4+n}\over\phi^n},\label{potential}
\end{equation}
where $\Lambda$ is a parameter with dimensions of mass
and the conformal factor increases with $\phi$, for example as
\begin{equation}
A(\phi) = \exp{(\beta\phi/M_p)},
\end{equation}
then the effective potential has a minimum that depends on the density.
Thus the physical properties of the field depend  on the ambient density.

From the above we can see that the mass of the chameleon is an increasing
function of the density such that it can be massive in dense environments,
but nearly massless cosmologically. Also, the non-linearity in the
chameleon equation of motion gives rise to the thin-shell mechanism,
which is discussed in the next section. { The  net effect of these properties
is that chameleon fields evade detection via gravitational tests for fifth forces, but can
 play the role of dark energy.}

\section{Thin-shells, Electrostatic Conductors, and the Chameleon Fifth Force}\label{thin_shell}

It is well-known from electrostatics that to find the backreaction of a charge {\blue on electric potential} close
to a conducting sphere  one may use the method of images.  Here we consider the analogous problem
for the chameleon mechanism, in that we consider the backreaction of a test
body close to a thin-shell body  that is a ``chameleon conductor" .  The method of images will
be effective insofar as the conductor approximation is valid, which in turn (as we argue) implies that the shell is
sufficiently thin.

{\blue Let us first summarize the chameleon properties close to a compact,
spherical body with a thin shell}.
The chameleon field inside a
body of radius $R_c$ with a thin shell is \cite{Khoury:2003rn}
\begin{eqnarray}
&&0<r<R_{\mathrm{roll}}:\qquad \phi\approx\phi_{c},\\
&&R_{\mathrm{roll}}<r<R_{c}:\qquad
\phi_{int}=\frac{\beta\rho_{c}}{3M_{p}}\left(\frac{r^{2}}{2}
+\frac{R_{\mathrm{roll}}^{3}}{r}\right)-\frac{\beta\rho_{c}R_{\mathrm{roll}}^{2}}{2M_{p}}+\phi_{c},
\end{eqnarray}
where $M_p=(8\pi G)^{-1/2}$, $\phi_c$ is the value of the field inside the
compact body of density $\rho_c$  and radius $R_c$, while $R_{roll}$ is the radius
where the field starts to move from its value inside the compact body. To make the
above two approximations have assumed:
\begin{eqnarray}
&&M_{p}\mid V_{,\phi}\mid\ll\beta\rho e^{\beta\phi/M_{p}}
\,\,\longrightarrow\,\, V_{,\phi} \,\,\mathrm{is\,\, negligible},\\
&&\beta\phi/M_{p}\ll1\,\,\longrightarrow\,\,e^{\beta\phi/M_{p}}\approx1.
\end{eqnarray}
We can write the field at the surface of the body, i.e.
$r=R_{c}$, as
\begin{equation}
\phi_s\equiv\phi(R_{c})=\frac{\beta\rho_{c}}{3M_{p}}\left(\frac{R_{c}^{2}}{2}
+\frac{R_{\mathrm{roll}}^{3}}{R_c}\right)-\frac{\beta\rho_{c}R_{\mathrm{roll}}^{2}}{2M_{p}}+\phi_{c}.\label{phi}
\end{equation}
If we are in the thin-shell regime, then
\begin{equation}\label{thinsh}
\frac{\Delta R_c}{R_c}\approx\frac{R_c-R_{\mathrm{roll}}}{R_c}\ll1
\end{equation}
and using a Taylor expansion,  eq. (\ref{phi}) reduces to
\begin{equation}
\phi_s-\phi_c=\frac{\beta\rho_c}{2M_p}(\Delta R_c)^2.\label{phis}
\end{equation}
Inserting the Newtonian potential
\begin{equation}
\Phi_N=\frac{1}{8\pi
M_p^2}\frac{M_c}{R_c}=\frac{\rho_cR_c^2}{6M_p^2}
\end{equation}
into  eq. (\ref{phis}) we obtain
\begin{equation}
\frac{\phi_s-\phi_c}{3\beta M_p\Phi_N}=\left(\frac{\Delta
R_c}{R_c}\right)^2,
\end{equation}
which provides a criterion for comparing the value of the field at the surface relative to its interior
in terms of the thickness of the shell. Whilst the above has been derived using
approximations, it has been checked analytically and numerically in the original
papers, both for $\beta=O(1)$ \cite{Khoury:2003rn} and in the strong
coupling regime \cite{Mota:2006fz}.
Using (\ref{thinsh}), we see that {\blue the field values
are nearly the same, i.e. the field is continuous across the thin shell.  Therefore, ignoring the
outside chameleon mass, the problem becomes analogous to that of a conductor held at fixed potential
in electrostatics, where $\phi$ takes a fixed value inside, $\phi_c$, while satisfying the Laplace
equation outside the object. As a result, the back-reaction problem can be treated analogously to that of a  conducting sphere in electrostatics.}


In obtaining the total force between a spherical
body with a thin-shell and a test  object without a
thin-shell,  we have three types of
forces to consider.

\begin{itemize}
\item The net gravitational force from a density distribution
$\rho(r)$ of the thin-shell body acting on a test object of mass
$m$.  This is given by
\begin{equation}
\vec{F}_G=-\frac{m}{M_p}\vec{\nabla}\phi_G(r),\label{FG}
\end{equation}
where $\phi_G$ is the gravitational potential satisfying:
\begin{equation}
\nabla^2\phi_G=\frac{\rho(r)}{2M_p}.\label{phiG}
\end{equation}

\item The chameleonic force acting on a test object of mass $m$ coupled
with strength $\beta$ to a chameleon field $\phi(r)$ of a
body with a thin-shell of mass $M$. Here the relevant force is \cite{Khoury:2003rn}
\begin{equation}
\vec{F}_{\phi}=-\frac{\beta}{M_p}m\vec{\nabla}\phi(r).\label{5force}
\end{equation}
Inserting the external solution for a thin-shell rigid body of
radius $R_c$
\begin{equation}
\phi_{ext}(r)=-\frac{\beta}{4\pi
M_p}\frac{\widetilde{M}e^{-m_{\infty}r}}{r}+\phi_{\infty},\quad\quad
r>R_c
\end{equation}
and ignoring the exponential factor (since
$m_{\infty}R_c\ll1$) the chameleon force (or fifth force) can be written as
\begin{equation}
\vec{F}_{\phi}=-\alpha\frac{m\widetilde{M}}{r^2}\hat{r}
\end{equation}
where $\alpha=\beta^2/4\pi M_p^2$ and $r$ is the distance between
the centers of  the bodies. We have introduced $\widetilde{M}$
as the reduced mass of the thin-shell body with  radius $R_c$ and
mass $M$
\begin{eqnarray}
\widetilde{M}=\frac{3\Delta R_c}{R_c}M,
\end{eqnarray}
where
\begin{eqnarray}
\frac{\Delta R_c}{R_c}=\frac{\phi_{\infty}-\phi_c}{6\beta
M_p\Phi_N},
\end{eqnarray}
where $\Phi_N=M_c/8\pi M_p^2 R_c$ is the Newtonian potential,
$\phi_{\infty}$ minimizes the effective potential outside the
thin-shell body and $\phi_c$ minimizes the effective potential
inside the thin-shell body which can be obtained as:
\begin{equation}
\phi_{\infty}=\left(\frac{\Lambda^5 M_p}{\beta
\rho_{\infty}}\right)^{1/2},\qquad\phi_i=\left(\frac{\Lambda^5
M_p}{\beta \rho_{c}}\right)^{1/2},
\end{equation}
where we have used (\ref{potential}) with $n=1$ in the above and
the density far away from the body with the thin-shell
is $\rho_{\infty}$, whilst the density inside the body is
$\rho_c$.

\item The back reaction force between the thin-shell body and test
object. As noted above, this can be obtained using method of images, yielding
\begin{equation}
\delta \vec{F}=-\frac{1}{8\pi
M_p^2}\frac{mm^{\prime}}{(r-r^{\prime})^{2}}\hat{r},\label{delF}
\end{equation}
where $m^{\prime}$ is the image mass of the test body located at
$r^{\prime}$ inside the thin-shell body of radius $R_c$:
\begin{equation}
m^{\prime}=-\frac{mR_c}{r},\qquad
r^{\prime}=\frac{R_c^2}{r}.\label{images}
\end{equation}
\end{itemize}

Note that we have neglected  the chameleon mass exterior to the
thin-shell body, since this mass  depends on the density of the
environment and the density outside the body is assumed to be
small {\blue (this will change in the next section, where we
assume a diffuse NFW profile)}. Accordingly, the total force
acting on the test body $m$ from the thin-shell body $M$,
including both the fifth force and back reaction, is
\begin{equation}
\vec{F}_T=\vec{F}_G+\vec{F}_{\phi}+\delta \vec{F}.
\end{equation}

 To summarize, the test object responds to not only the
fifth force from the thin shell, but also to its own image
reflected from the thin shell. While the primary fifth force is
attractive, the back-reaction (or self-force) is repulsive,
as the image has the opposite charge (i.e. negative mass). Moreover,
the self-force will become comparable to the primary force as test object approaches the thin shell.

\section{NFW profile and chameleon force}\label{NFW}

In this section we will obtain the chameleon force for a dark matter halo (similar to that of the Milky
Way galaxy) acting on a test body  by considering the
NFW profile for the halo density \cite{Navarro:1996gj}, which is of the form
\begin{equation}
\rho(r)=\frac{\rho_s}{\frac{r}{r_s}(1+\frac{r}{r_s})^2} \label{rho}
\end{equation}
where $r_s =
10\,\mathrm{kpc}=1.57\times10^{27}\,\mathrm{eV}^{-1}$. The
quantity
 $\rho_s$ is given by
\begin{equation}
\int^{300\,\mathrm{kpc}}_{0}4\pi
r^2\rho(r)dr=200 \bar{\rho}_c \left[\frac{4\pi}{3}(300\,\mathrm{kpc})^3\right]
\end{equation}
where $\bar{\rho}_c=3H_0^2M_p^2$ is the citical density of the
universe. Using the present Hubble constant  $H_0 \simeq 71
~{\rm km/s/Mpc} = 1.5 \times 10^{-33}\,\mathrm{eV}$ and the
reduced Planck mass $M_p= (8\pi G)^{-1/2} = 2.43\times
10^{27}\,\mathrm{eV}$ we obtain
$$
\rho_s=3.0\times 10^{-5}\,\mathrm{eV}^4.
$$
We also consider the potential as in Eq. (\ref{potential}).

Next we divide the space into two distinct regions.

\textbf{I)} The interior region. Here the following constraint
\begin{equation}
\frac{M_p\nabla^2\phi}{\beta\rho(r)}\ll1\label{consint}
\end{equation}
is satisfied, and so
  the chameleon equation becomes
\begin{equation}
\nabla^2\phi=V_{,\phi}+\frac{\beta\rho(r)}{M_p}e^{\beta\phi/M_p} .
\end{equation}
Neglecting $\nabla^2\phi$ and solving
\begin{equation}
\frac{\beta\rho(r)}{M_p}+V_{,\phi}=0\label{intex}
\end{equation}
where the exponential coefficient has been ignored since we have
assumed $\beta\phi/M_p\ll1$, we find
\begin{equation}
\phi_{int}(x)\approx \phi_s\left[x(1+x)^2\right]^{1/(n+1)},\qquad
x_{0{\small <}}<x<x_{0{\small >}}\label{inte}
\end{equation}
for the solution to eq. (\ref{intex}).
For convenience we have defined the dimensionless variable
$x=r/r_s$ and
\begin{equation}
\phi_s=\left(\frac{n\Lambda^{n+4}
M_p}{\beta\rho_s}\right)^{1/(n+1)}.
\end{equation}
The quantities $x_{0<}$ and $x_{0>}$ respectively denote the lower and upper
values of $x$ that beyond which the constraint (\ref{consint}) is no
longer valid. We can find these values by setting
$\frac{M_p\nabla^2\phi}{\beta\rho}=0.01$, and then solving
numerically for the two real roots $x_{0<}$ and $x_{0>}$.

\begin{figure}[tbp]
\centering {\includegraphics[width=.5\textwidth]{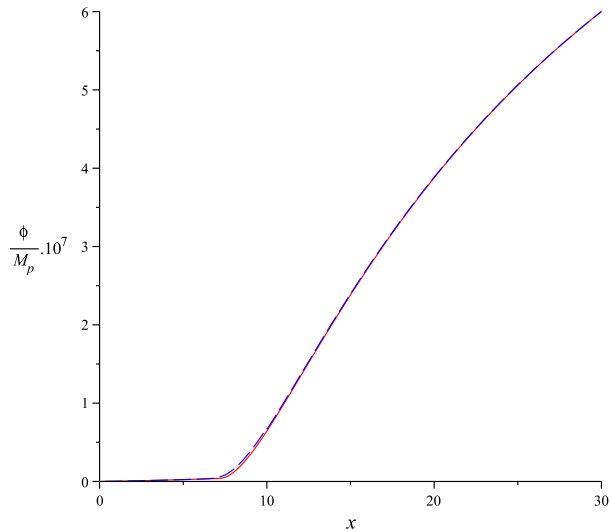}}
\caption{Chameleon field $\phi$ versus distance $x=r/r_s$, where the red
solid line shows the analytic solution and the blue dashed one depicts
the numeric solution.} \label{px}
\end{figure}

\textbf{II)} The exterior region. Here we impose the condition
$$
V_{,\phi}\ll\nabla^2\phi
$$
which we take to be valid for  $x>x_c$.
The exterior solution can therefore be found by  neglecting the potential term in the chameleon equation
\begin{equation}
\nabla^2\phi_{ext}(x)=\frac{\beta r_s^2\rho(x)}{M_p},
\end{equation}
yielding
\begin{equation}
\phi_{ext}(x)\approx-\frac{B\ln(1+x)}{x}-\frac{C}{x}+\phi_{\infty},\qquad
x>x_c\label{exte}
\end{equation}
with
\begin{eqnarray}
B=\frac{\beta\rho_sr_s^2}{M_p},
\end{eqnarray}
where $C$ and $\phi_{\infty}$ are integration constants.

The quantity $\phi_{\infty}$ can be interpreted as the minimum of
chameleon field for $x\gg x_c$.  However, we need some criterion for
determining $C$ and $x_c$.  Recalling that at $x=x_{0>}$ the chameleon field starts
to deviate from the interior constraint (\ref{consint})  and that the exterior solution is valid
for $x\geq x_c$, we find that  if we demand
\begin{equation}
\frac{x_c-x_{0>}}{x_c}\ll1\label{thinshell}
\end{equation}
then to a very good degree of approximation the interior solution
is still valid inside the shell $x_{0>}<x<x_c$, and the shell can
be considered as sufficiently thin. Hence to compute the
integration constant $C$ we  match the interior and exterior
solutions and their derivatives at $x=x_c$, which yields
\begin{eqnarray}
C=-B\ln{(1+x_c)}+\phi_{\infty}x_c-\phi_s\left[x_c(1+x_c)^2\right]^{1/(n+1)}x_c,\label{eqC}
\end{eqnarray}
where $x_c$ is the smaller real root of the following equation
\begin{eqnarray}
\left[x_c(1+x_c)^2\right]^{1/(n+1)}=\frac{(n+1)}{\phi_s}\,\frac{(1+x_c)\phi_{\infty}-B}{(n+4)x_c+n+2} \label{eqXc}
\end{eqnarray}
which can be solved numerically for given parameters $n$,
$\beta$, $\Lambda$ and $\phi_{\infty}$.

For simplicity we henceforth set $n=1$ and take
$\beta$ of order unity. Whilst we could work with different parameters and
in the strong coupling regime, we would not expect our results to change
qualitatively, as long as the thin-shell conditions are satisfied.
Then the only parameters left to be determined are $\phi_{\infty}$
and $\Lambda$. We determine these via  two basic considerations.
{First, the thin-shell conditions (\ref{consint}) and
(\ref{thinshell}) must be satisfied, where the former yields an
upper bound for $\Lambda$ of the order of $10\,\mathrm{eV}$.
Second, since the size of the halo is approximately
$300\,\mathrm{kpc}$ then  the values for $\phi_{\infty}$ and
$\Lambda$ must imply that the smaller real root of eq.
(\ref{eqXc}) be in a range between $0<R_c<300\,\mathrm{kpc}$,
where $R_c=(10~\,\mathrm{kpc})x_c$. Numerically we find that
the second consideration along with the upper bound on $\Lambda$
imply the upper bound for $\phi_{\infty}$ is of the order of
$10^{-5}M_p$. Hence we choose $\phi_{\infty}=1.5\times10^{-6}M_p$
and $\Lambda=4.66\,\mathrm{eV}$, which satisfies all
considerations simultaneously,  yielding $x_c=7.183$ \footnote{Note that this form of the potential, $V(\phi)$, is only assumed on large galactic scales, and thus cannot be directly compared to solar system or cosmological constraints.} .}
\begin{figure}[tbp]
\centering {\includegraphics[width=.5\textwidth]{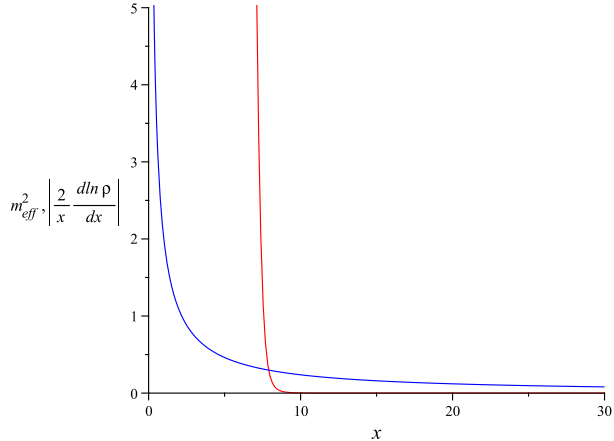}}
\caption{The effective mass of the chameleon field (red) and
$|\frac{2}{x}\frac{d\ln{\rho}}{dx}|$ (blue) versus distant
$x=r/r_s$. It is clear that for the internal region $x<x_c=7.183$ we
have $m_{eff}^2\gg|\frac{2}{x}\frac{d\ln{\rho}}{dx}|.$}
\label{masseff}
\end{figure}

We now proceed to solve the whole chameleon equation numerically,
using the information from our semi-analytic investigation thus
far. Specifically, setting $n=1$, $\beta=1$ and
$\Lambda=4.66\,\mathrm{eV}$ yields
\begin{equation}
\frac{1}{x^2}\frac{d}{dx}\left(x^2\frac{d\widetilde{\phi}(x)}{dx}\right)=
\frac{\beta\rho_sr_s^2}{M_p^2x(1+x)^2}e^{\beta\widetilde{\phi}(x)}-\frac{n\Lambda^{n+4}
r_s^2}{M_p^{n+2}\widetilde{\phi}(x)^{n+1}}
\end{equation}
while we have re-scaled $\phi/M_p\rightarrow \widetilde{\phi}$.  We present our numerical and semi-analytical solutions in
fig. (\ref{px}) .

It is evident that the numerical and semi-analytic solutions
match with high accuracy, confirming the approximations we made for this choice of parameters.
For example, one approximation neglected the mass of the chameleon
field in the external region. Fig. (\ref{masseff}) shows the
effective mass of the  chameleon, i.e.
$m_{eff}^2=d^2V_{eff}(\phi)/d\phi^2$, versus distant $x$. Here we
see that the chameleon mass almost vanishes for $x>x_c$, and that
$m_{eff}^2\gg|\frac{2}{x}\frac{d\ln{\rho}}{dx}|$ for the interior
region. {The latter inequality is an equivalent expression for the
interior constraint (\ref{consint}): taking the derivate with respect to
$r$ from Eq. (\ref{intex}) gives
$m_{eff}^2=-\beta\rho^{\prime}/\phi^{\prime}M_p$ where $\phi$ is
the interior solution given by eq. (\ref{inte}).}

In the next section, we  use our semi-analytic solutions
(\ref{inte}) and (\ref{exte}), to obtain physical quantities such as the fifth force, its back-reaction, and their possible observational consequences.

\section{circular velocity from fifth force with back reaction}\label{circ_vel}

In this section,  as an application of the thin-shell framework developed above, we will obtain the circular velocity of
orbiting satellite galaxies in the exterior region of the galactic haloes, caused by the
gravitational force, the fifth force and the back-reaction. Since by the constraint (\ref{thinshell}) we
demanded the shell of the halo to be thin,  we will use the
method of images which taks into account the back-reaction
effect of the satellites.

Furthermore, for simplicity, we limit ourselves to
those satellites in the galaxy that can reasonably be considered as
bodies {\it without} a thin-shell. For these satellites we have
\begin{equation}
\frac{\phi_{ext}(x)-\phi_c}{6\beta M_p \Phi_N}>1,\label{noshelsat}
\end{equation}
where $\phi_{ext}(x)$ is the external chameleon field (\ref{exte})
calculated at the satellite location. Since the virial
velocity of a satellite is $v_{vir}^2\propto\Phi_N$, the
condition   (\ref{noshelsat}) yields a
constraint on the virial velocity
\begin{equation}
v_{\rm vir}^2<\frac{\phi_{ext}(x)}{6\beta M_p}
\end{equation}
where we have neglected $\phi_c$, which is the minimum of the
chameleon field inside the rigid body. For our choice of NFW halo and coupling parameters, this reduces to $ v_{\rm vir} \lesssim $ 10-15 km/s,
close to the thin shell, which is a typical velocity dispersion for intermediate mass satellites of Milky Way.

Inserting the NFW profile (\ref{rho}) for the density distribution
of the halo in the eq. (\ref{FG}), the net gravitational force
(that is finite at the origin) is
\begin{equation}
\vec{F}_G=-\frac{m}{M_pr_s}\left[-\frac
{\widetilde{B}}{x(1+x)}+\frac{\widetilde{B}\ln(1+x)}{x^2}\right]\hat{r}\label{FGh}
\end{equation}
where $\widetilde{B}=\rho_sr_s^2/2M_p$.

The fifth force acting on a test object of mass $m$ caused by
the thin-shell halo can be obtained from eqs. (\ref{5force}) and
(\ref{inte}) for the interior region as
\begin{equation}
\vec{F}_{\phi_{int}}=-\frac{\beta m}{M_pr_s}\left[\frac
{1+3x}{x(1+x)(1+n)}\phi_{int}\right]\hat{r}\qquad\quad
x<x_c\label{Cin5force}
\end{equation}
or from (\ref{exte})
\begin{equation}
\vec{F}_{\phi_{ext}}=-\frac{\beta m}{M_pr_s}\left[\frac
{C}{x^2}-\frac {B}{x(1+x)}+\frac{B\ln(1+x)}{x^2}\right]\hat{r}
\qquad\quad x>x_c\label{Cex5force}
\end{equation}
for the exterior region.

The force caused by the back-reaction effect is simply the radial force between the
test object $m$   and its image mass $m^{\prime}$, located at $r^{\prime}$ inside the
thin-shell halo.  This  is given by eqs. (\ref{delF}) and
(\ref{images}) as
\begin{equation}
\delta\vec{F}=\left(\frac{m^2}{8\pi
M_p^2r_s^2}\right)\frac{x_cx}{(x^2-x_c^2)^2}\hat{r}\label{CdelF}
\end{equation}
where we have used the dimensionless radial coordinate $x=r/r_s$
and $x_c$ is the smaller real root in eq. (\ref{eqXc}). Finally,
the total force is
\begin{equation}
\vec{F}_{T}=\vec{F}_G+\vec{F}_{\phi}+\delta\vec{F},
\end{equation}
which is the sum of the net gravitational
force (\ref{FGh}), the fifth force
(\ref{Cin5force})/(\ref{Cex5force}) and the image force (\ref{CdelF}).
\begin{figure}[tbp]
\centering {\includegraphics[width=.5\textwidth]{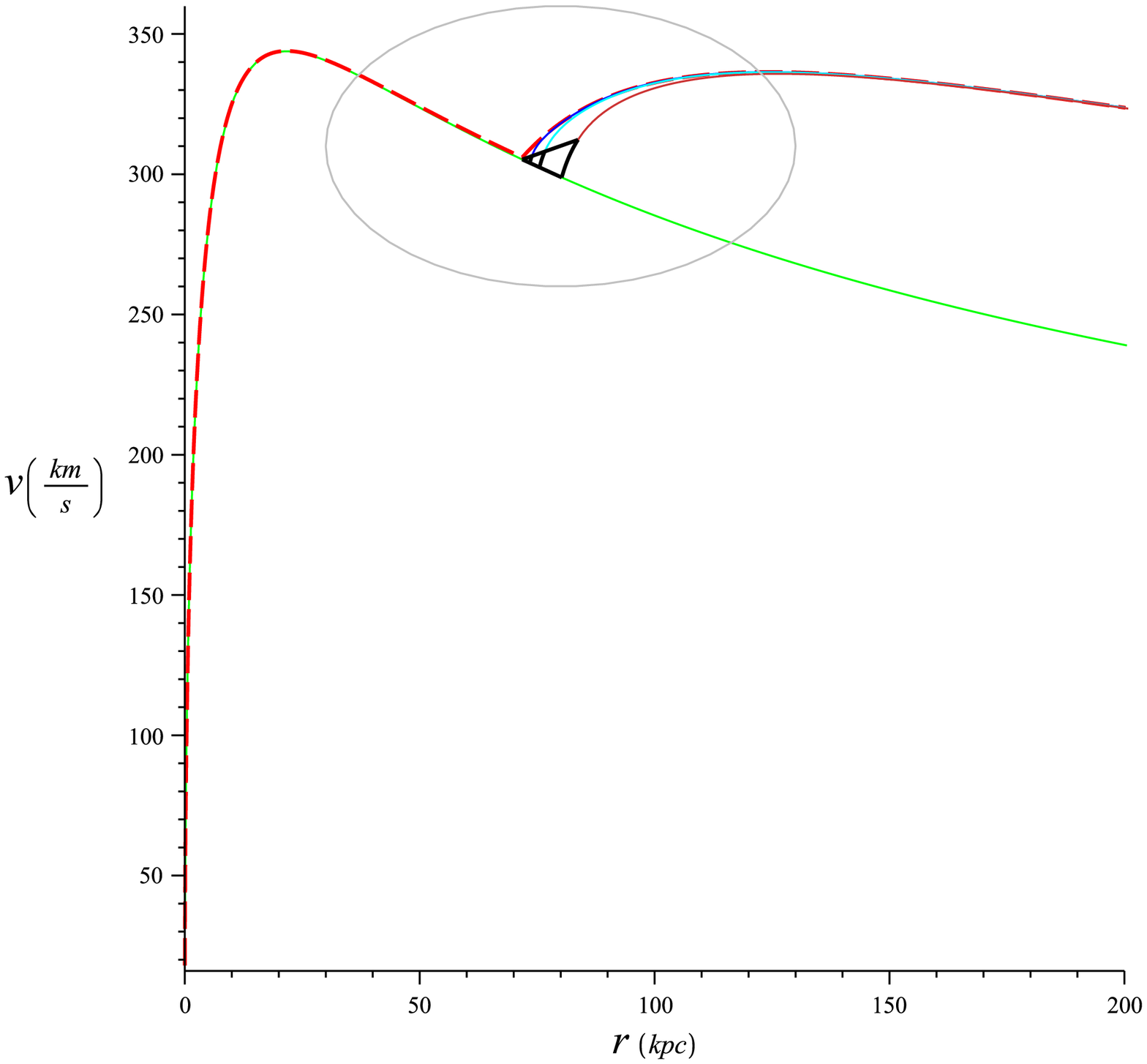}\,\,}
\centering
{\includegraphics[width=.4\textwidth]{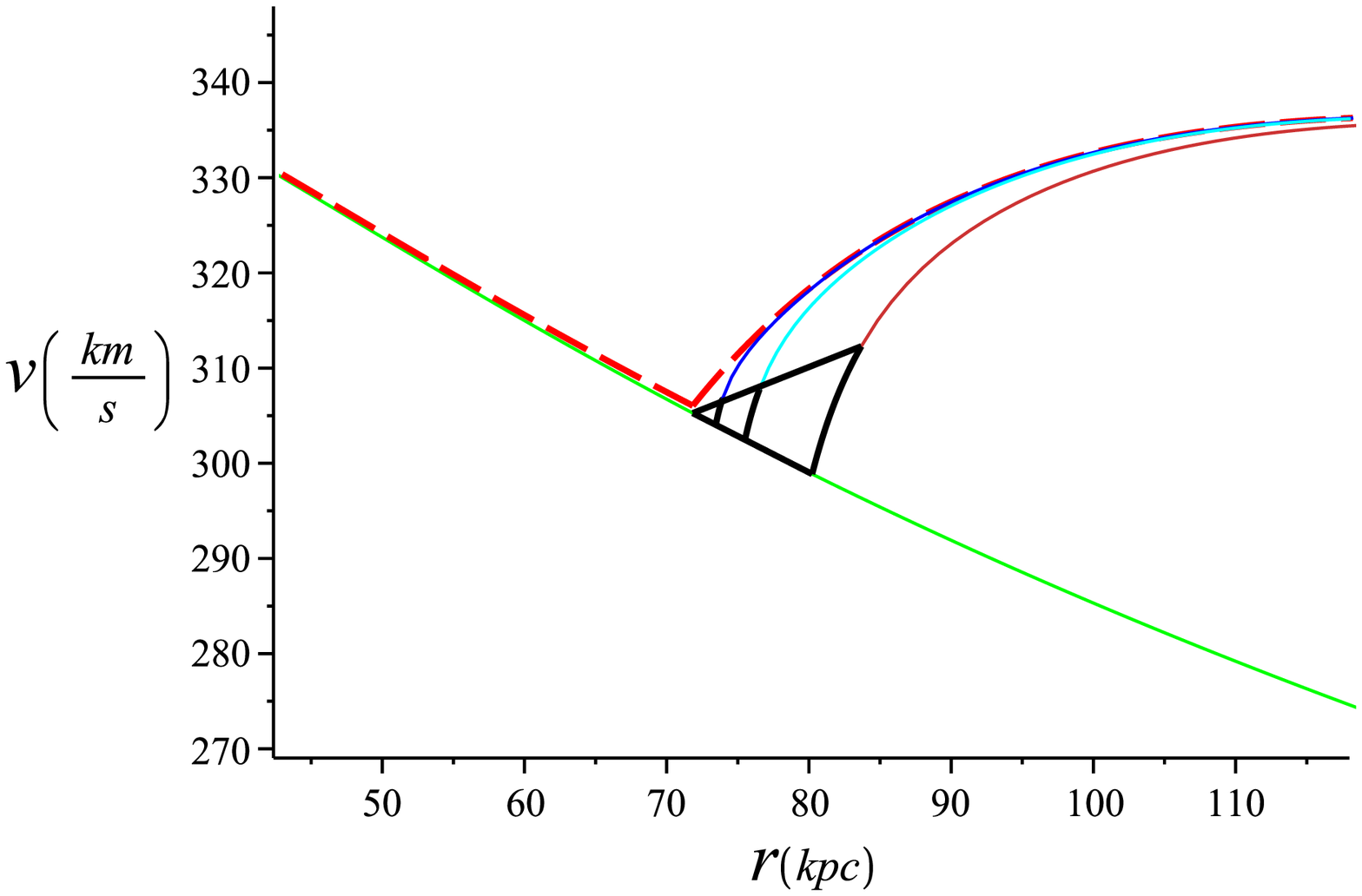}\,\,}
\caption{Circular velocity versus distance from the center of the galactic halo. The green solid line is obtained by considering the net
gravitational force while the red dashed one is obtained by
considering the correction from including the fifth force. For the rest, the
correction due to the back-reaction has been also taken into account. The
mass of the test object increases from top to bottom:
$m=10^{8}M_{\odot}$ (blue), $m=10^{9}M_{\odot}$ (cyan),
$m=10^{10}M_{\odot}$ (brown). } \label{vel}
\end{figure}

From the radial force we  may obtain the circular velocity
$v(r)=\sqrt{|\vec{F}|r/m}$ as a function of $r$,  the distance
from center of the halo.
Including only
the fifth force, it is easy to see that  the circular
velocity is the same for all the objects in the external region of
the halo, i.e. it is independent of the mass of the test object.
This changes once the back-reaction is taken into account, which is shown in Fig. (\ref{vel}). For
example, we see that for an object with mass of the order $10^{10} M_{\odot}$,
considering just gravitational force from NFW profile we will
obtain the green line for the circular velocity. If we
include the fifth force from the chameleon theory for the NFW profile we
get the red dashed line, which has a negligible effect in the
interior region, but manifestly changes the exterior behaviour. Including next
the back-reaction we obtain the brown
line, which clearly deviates from the dashed red one for a large
satellite mass. For smaller masses the deviation is not as big: in fact,
it is clear that the correction from the image force becomes
less important as the mass of the satellite deceases. The
region enclosed by the black triangle denotes the regime in
which the back-reaction becomes $50\%-100\%$ of the original fifth-force. Therefore, we cannot trust our approximations
 (which neglected the effect of satellite on the thin shell); the velocity should lie somewhere inside the triangle,
  which specifies the range of uncertainty in our prediction. This triangle region grows as the mass of the test body
increases, a feature more vivid in the close-up.

An interesting observational window into violations of equivalence principle on Galactic scales was introduced
in \cite{Kesden:2006zb,Kesden:2006vz,Kesden:2009bb}, where it was argued that leading and trailing tidal streams
of satellites  of Milky Way would be asymmetric, if stars and dark matter (that dominates these galaxies) experienced
different gravitational accelerations.
 Based on the symmetry of the tidal streams of the Sagittarius dwarf galaxy, \cite{Kesden:2006zb}
 argue that the difference between these accelerations should be $< 10\%$, which implies $<5\%$ difference
 in circular velocities. Given that $v_{\rm vir} \sim 15$ km/s, and $M \sim 10^9 M_{\odot}$ for Sagittarius dwarf
 \cite{Lokas:2010wu}, we expect around $3\%$ difference in circular velocities (see
Fig. \ref{vel}),
 which  is just below the observational limit, assuming that Sagittarius orbit is just outside a chameleon thin shell.
 However, further improvements in these limits, and/or study of other tidal streams in the Milky Way halo could well provide
 a way to discover this effect in the outer Galaxy.

\section{Conclusions}\label{conclude}
In chameleon theories there is a fifth force, which is suppressed
due to the chameleonic effect and the thin shell. In this
work, for the first time, we have quantified the proper thin shell conditions in chameleon gravity for the dark matter haloes surrounding galaxies (such as Milky Way), assuming an NFW density profile. As a result we could obtain the chameleon force for such a profile, which adds to the net
gravitational force. We showed that this additional fifth force manifestly changes the behaviour of the circular velocity for the
test objects in the exterior region of the halo.

Since chameleon gravity is non-linear, the addition of a test object can change the chameleon field, and thus lead to a back-reaction or self-force, an effect that has never been calculated explicitly. In this paper, we have
calculated this back-reaction, or self-force, in chameleon
theories on galactic scale objects. Our method used the analogy
between gravitational objects with a thin shell and
electrostatics, enabling us to use the methods of images to
compute the back-reaction. We applied our method to the NFW
profile of dark matter halos.

When we apply our results to the circular velocity of satellites in the surrounding region of the halo we have found that the back-reaction cannot be ignored.
When we only included the fifth force we found that the circular velocity
was the same for all objects in the exterior region surrounding the halo.
This situation changes considerably upon including the back-reaction.
Indeed, depending on the mass of the satellite, the back-reaction can be a large modification
to the original result.

Our results suggest that there could be a violation of
the equivalence principle in the outskirts of galactic haloes. While current bounds, based on the observations of the tidal streams of the Sagittarius dwarf galaxy, are not yet sensitive to this effect, future surveys of kinematics in the Milky Way halo can dramatically strengthen these limits \cite{Kesden:2006vz}.    This opens up
another channel for testing chameleon theories. More generally our results suggest that
in some of the gravitational tests for chameleon theories the back-reaction
should be taken into account when putting constraints on the parameters of
the theory. This is currently in progress.

\section*{Acknowledgment}
We wish to thank Justin Khoury, Louie Strigari, and Philippe Brax for discussions and useful comments. This
work was supported in part by the Natural Sciences and Engineering Research
Council of Canada and by STFC, UK. NA is in part supported by the
Perimeter Institute for Theoretical Physics. Research at Perimeter
Institute is supported by the Government of Canada through Industry
Canada and by the Province of Ontario through the Ministry of Research
\& Innovation. ACD wishes to thank the Perimeter Institute
for hospitality whilst this work was initiated.

\end{document}